\documentclass[sigconf]{acmart}
\acmConference[KDD-MLF 2026] {KDD 9th Workshop on Machine Learning in Finance}{August 9--13, 2026}{Jeju Island, Republic of Korea.}
\AtBeginDocument{%
  }

\setcopyright{acmlicensed}
\copyrightyear{2026}
\acmYear{2026}
\acmPrice{}
\acmISBN{}
\acmDOI{}

\usepackage{xspace}
\usepackage{multicol, multirow, threeparttable}
\usepackage{graphicx}
\usepackage{caption}
\usepackage{subcaption}

\usepackage{fontawesome}
\usepackage{enumitem}

\newcommand{\name}{FinMamba\xspace}
\usepackage[capitalize]{cleveref}
\crefname{section}{Sec.}{Secs.}
\Crefname{section}{Section}{Sections}
\Crefname{table}{Table}{Tables}
\crefname{table}{Tab.}{Tabs.}
\crefformat{equation}{Eq.~#2#1#3}

\usepackage{pifont} 
\usepackage{bbding}
\usepackage{fontawesome}

\definecolor{mypurple}{rgb}{0.7, 0, 0.9}
\definecolor{myorange}{rgb}{0.9, 0.7, 0}
\definecolor{bl}{rgb}{0.25, 0.5, 0.9}
\newcommand{\best}[1]{{\textbf{\textcolor{red}{#1}}}}
\newcommand{\second}[1]{{\textcolor{bl}{\underline{#1}}}}

\usepackage{fancyhdr}
\begin{document}

\title{\name: Market-Aware Graph Enhanced Multi-Level Mamba for Stock Movement Prediction}
\renewcommand{\shorttitle}{\name: Market-Aware Graph Enhanced Multi-Level Mamba for Stock Movement Prediction}

\author{
    Yifan Hu\textsuperscript{\rm 1,}*, 
    Peiyuan Liu\textsuperscript{\rm 1,}*, 
    Yuante Li\textsuperscript{\rm 3}, 
    Dawei Cheng\textsuperscript{\rm 2},\\
    Naiqi Li\textsuperscript{\rm 1}, 
    Tao Dai\textsuperscript{\rm 4,}\footnotemark[2],
    Jigang Bao\textsuperscript{\rm 1}, 
    Shu-tao Xia\textsuperscript{\rm 1}
}
\affiliation{
    \textsuperscript{\rm 1}Tsinghua University \ \ \ \ 
    \textsuperscript{\rm 2}Tongji University \ \ \ \ 
    \textsuperscript{\rm 3}Carnegie Mellon University \ \ \ \ 
    \textsuperscript{\rm 4}Shenzhen University\\
    \textsuperscript{*}Equal Contribution \ \ \ \ 
    \textsuperscript{\footnotemark[2]}Corresponding Author\\
    \faEnvelopeO\ Primary contact: huyf0122@gmail.com\\
    \country{}
}

\renewcommand{\shortauthors}{Yifan Hu, Peiyuan Liu, Yuante Li, Dawei Cheng, Naiqi Li, Tao Dai, Jigang Bao, Shu-tao Xia}

\begin{abstract}
  Recently, combining stock features with inter-stock correlations has become a common and effective approach for stock movement prediction. However, financial data presents significant challenges due to its low signal-to-noise ratio and the dynamic complexity of the market, which give rise to two key limitations in existing methods. First, the relationships between stocks are highly influenced by multifaceted factors including macroeconomic market dynamics, and current models fail to adaptively capture these evolving interactions under specific market conditions. Second, for the accuracy and timeliness required by real-world trading, existing financial data mining methods struggle to extract beneficial pattern-oriented dependencies from long historical data while maintaining high efficiency and low memory consumption. 
  To address the limitations, we propose \name, a Mamba-GNN-based framework for market-aware and multi-level hybrid stock movement prediction.
  Specifically, we devise a dynamic graph to learn the changing representations of inter-stock relationships by integrating a pruning module that adapts to market trends. Afterward, with a selective mechanism, the multi-level Mamba discards irrelevant information and resets states to skillfully recall historical patterns across multiple time scales with linear time costs, which are then jointly optimized for reliable prediction. 
  Extensive experiments on U.S. and Chinese stock markets demonstrate the effectiveness of our proposed \name, achieving state-of-the-art prediction accuracy and trading profitability, while maintaining low computational complexity. The code is available at \url{https://github.com/TROUBADOUR000/FinMamba}.
\end{abstract}




\maketitle

\section{Introduction}

Stock movement prediction plays a pivotal role in data science-driven quantitative trading applications due to its potential to guide profitable investment strategies~\cite{survey,cisthpan,MDGNN}. Unlike general time series forecasting~\cite{hu2025bridging,hu2025timefilter}, stock prices record human-brain-armed game behaviors, which are influenced by various factors, including investor behavior, economic indicators, political events, and global news. 
This high level of volatility and the multifaceted nature of these influences make accurate prediction a particularly daunting challenge. 
Traditional machine learning approaches~\cite{alpha,randomforest} have been used to address these challenges, but often require expert-designed features and struggle to capture the intricate dependencies between stocks. Recently, deep learning-based methods~\cite{CTTS,ALSP,lsrigru} have shown great promise in overcoming these limitations by effectively combining stock features with inter-stock correlations~\cite{tcgpn}. However, such a paradigm could still be unsatisfactory due to the following two limitations. 

\begin{figure}[t]
    \centering
    \includegraphics[width=0.45\textwidth]{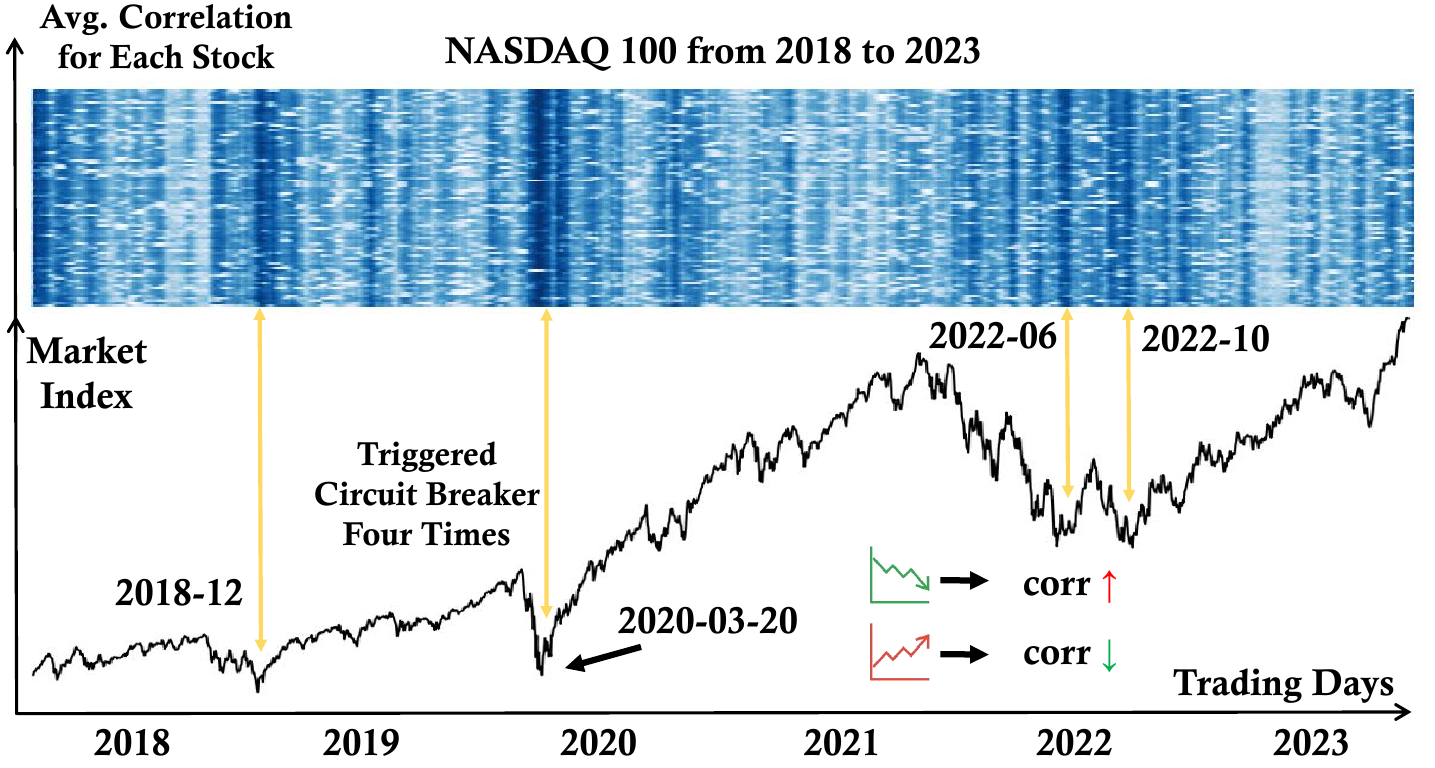}
    \caption{
         Market fluctuations have affected the correlation among stocks. In the heatmap above, the horizontal axis represents trading days, while the vertical axis displays the average correlation of each stock with other stocks on a given day. The line chart below reflects the macroeconomic market volatility. The analysis reveals that, during past market downturns, stock correlations tend to intensify. A discernible pattern emerges: stock correlations increase when the market index falls and diminish when the market index rises.
    }
    \Description{..}
    \label{fig:corr&index}
\end{figure}

\ding{182} \textbf{\textit{Multifaceted inter-stock relationships under various market conditions.}}
As is well known, synergy within the stock market is a key aspect, with related stocks often moving in sync~\cite{ALSP}, providing valuable insights for more accurate predictions. Meanwhile, inter-stock relationships are influenced by various factors, such as sector dynamics, regulatory politics, and the macroeconomic market environment. Therefore, comprehensive modeling of relationships between stocks is crucial.
Some of the existing methods~\cite{StockFormer,hist,ALSP} rely on incomplete prior knowledge, which may introduce biases. For instance, solely grouping stocks by industry sectors can overlook cross-sector influences from broader macroeconomic factors. 
Additionally, stock relationships are dynamic, evolving with market fluctuations, making it difficult for static models to accurately reflect these changes. In contrast, posterior-based methods construct adjacency matrices to model dynamic correlations between stocks based on dynamic time warping~\cite{cisthpan}, Pearson correlation coefficients~\cite{thgnn}, Euclidean distance~\cite{tcgpn} or learning-based methods~\cite{ADBTR,master}. However, posterior methods can lead to spurious correlations, where stocks with no actual relationship exhibit high sequence similarity due to coincidental trends. Thus, we propose a graph-based structure to integrate both static prior correlations, such as industry classifications, and dynamic posterior correlations derived from stock price sequences, providing a more comprehensive framework for capturing inter-stock dependencies.

Furthermore, the role of macroeconomic markets in shaping stock relationships is often overlooked, despite their significant influence on inter-stock correlations~\cite{market}. As the stock market evolves, the inter-stock relationships also shift dynamically. As \cref{fig:corr&index} depicts, during financial crises, stocks within the same industry tend to exhibit stronger correlations due to shared risk factors and similar investor behavior~\cite{DIMPFL201410}. This phenomenon is amplified by the impact of market indices, which heighten the influence of macroeconomic conditions on stock performance during turbulent periods~\cite{MA2022102940}. Therefore, market indices are not only critical for gauging overall market performance but also play a key role in driving the dynamic relationships between stocks. To better capture these shifts, we propose a graph optimization strategy that refines stock relationship modeling by leveraging market index feedback, retaining dominant edges based on index movements while pruning those that fail to reflect dynamic changes. This approach enables the model to adaptively adjust inter-stock correlations in response to market fluctuations, resulting in more flexible predictions.

\begin{figure}[t]
    \centering
    \includegraphics[width=0.475\textwidth]{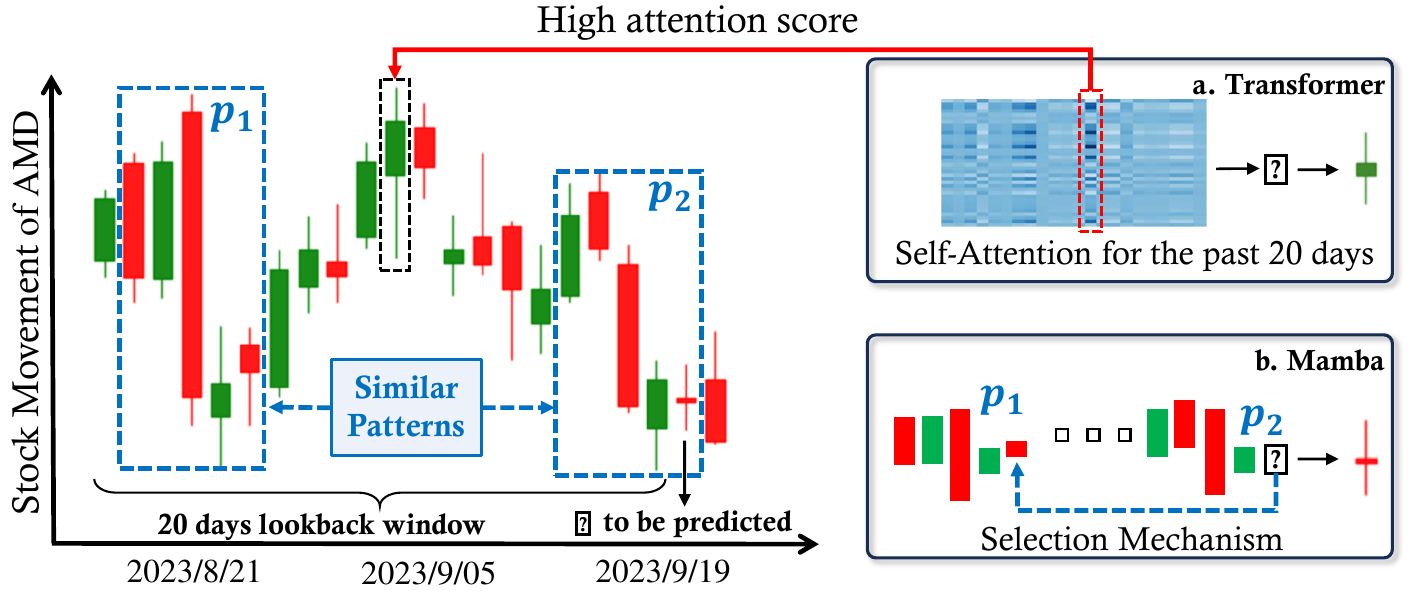}
    \caption{
    Comparison of (a) Transformer's self-attention mechanism and (b) Mamba's selective mechanism in stock movement prediction. Mamba is more effective in recalling similar historical patterns and avoiding overemphasis on outliers compared to Transformer.
    }
    \Description{..}
    \label{fig:intro}
\end{figure}


\ding{183} \textbf{\textit{Pattern-oriented dependencies under the timeliness constraints.}}
The time series studied in the deep learning literature~\cite{pdf,timebridge,hu2026existence} tend to exhibit some regularity (e.g., solar energy, traffic flow), facilitating the extraction of highly relevant features from similar repetitive patterns~\cite{generative}.
In contrast, stock prices often lack regular repetitive patterns due to low signal-to-noise ratios~\cite{FactorVAE}, with similar patterns appearing at different periods and amplitudes.
Recently, the Transformer~\cite{attention}, which calculates interactions between all sequence elements to capture their dependencies, has become a mainstream approach.
However, its self-attention mechanism often assigns disproportionate importance to outliers, leading to an overemphasis on anomalous patterns and hindering the effective capture of smoother, more common trends across different time scales~\cite{amd} (see \cref{fig:intro}(a)). 
After that, Mamba~\cite{mamba} has shown great potential in dynamic sequence modeling, with notable success in the other fields~\cite{mambair, mambavision}. The selective mechanism of Mamba enables it to recall temporal patterns from historical inputs, making the model particularly well-suited for capturing temporal dependencies in stock prices (see \cref{fig:intro}(b)). 
Moreover, real-time market trading imposes strict efficiency requirements for timely decision-making~\cite{ccso}. In light of this, Mamba offers a key advantage with its lower linear complexity compared to the Transformer, significantly enhancing prediction efficiency.

Additionally, stock prices exhibit distinct characteristics across different time levels, gradually transitioning from micro to macro levels~\cite{LUO2021101512}. On short-term, minute-level levels, price fluctuations are primarily driven by market microstructures, such as trader sentiment, high-frequency trading strategies, and market liquidity, often influenced by news, rumors, or technical indicators. As the time horizon extends to daily levels, mid-term factors like corporate performance, industry trends, and policy changes come into play~\cite{LU201777}. At weekly or monthly levels, macroeconomic forces such as economic growth, inflation, and monetary policy become the dominant drivers, reflecting structural market shifts and long-term trends. Therefore, analyzing stock price movements on a single time scale fails to comprehensively capture the full spectrum of temporal dependencies~\cite{ijcai2020p640}.

Based on the above motivations, we propose \name with Market-Aware Graph (MAG) and Multi-Level Mamba (MLM) to address existing limitations on the 1) multifaceted inter-stock relationships and 2) pattern-oriented characteristics of stock prices. 
Specifically, the MAG module adapts to effectively capture inter-stock relationships by integrating both the posterior and prior information, while refining according to macroeconomic market conditions. Then, the MLM module jointly models the micro- and macro-time level dependencies across the financial time series. Extensive experiments on U.S. and Chinese stock markets demonstrate the effectiveness of the proposed \name, along with its low computational complexity. 
In a nutshell, our contributions are summarized as follows:

\begin{itemize}[itemsep=-0.11em]
    \item To the best of our knowledge, this is the first work that enhances the inter-stock relationships under specific macroeconomic market conditions, leading to a more impressive understanding of market dynamics.
    \item We propose \name that constructs a Market-Aware Graph integrating both static priors and dynamic posteriors, refines these relationships using market index feedback, and models stock pattern-oriented dependencies across multiple levels through a Multi-Level Mamba framework.
    \item Extensive experiments on both U.S. and Chinese real-time stock markets demonstrate the state-of-the-art performance of the proposed \name with high computational efficiency. The visualization results provided valuable insights into the dynamics of stock patterns.
\end{itemize}



\section{Related Work}

\subsection{Stock Movement Prediction}
Due to the growing interest in stock market investing and the evolution of deep learning, most contemporary research endeavors now center around implementing and refining deep learning models in the realm of stock movement prediction. 
Consequently, combining stock features with inter-stock correlations has become a common and effective approach for stock movement prediction.
For example, THGNN~\cite{thgnn} employs a temporal and heterogeneous graph framework to extract insights from price history and relationships. MASTER~\cite{master} mines the momentary and cross-time stock correlation with learning-based methods. CI-STHPAN~\cite{cisthpan} constructs channel independent hypergraphs among stocks with similar stock price trends based on dynamic time warping.
Despite their success, current methodologies may still fall short by overlooking the fact that the relationship between individual stocks and the market index is such that it strengthens when the index falls and weakens when the index rises, especially during periods of market volatility.

\subsection{Mamba and State Space Models}
State Space Models (SSMs)~\cite{s4,mamba,mamba2} have emerged as a promising architecture for sequence modeling. Mamba, leveraging selective SSMs and a hardware-optimized algorithm, has achieved strong performance in several areas, including Natural Language Processing~\cite{mamba}, Computer Vision~\cite{mambair, mambavision}. 
For time series forecasting, Timemachine~\cite{timemachine} selects contents
for prediction against global and local contextual information with four SSM modules. Chimera~\cite{chimera} incorporates 2-dimensional SSMs with different discretization processes and input-dependent parameters to dynamically model the dependencies.
However, the application of Mamba in quantitative trading is still in its infancy, with approaches such as MambaStock~\cite{mambastock} applying the original Mamba for individual stock modeling and SAMBA~\cite{samba} using bidirectional Mamba blocks to capture long-term dependencies.
In our work, we enhance Mamba by modeling similar stock patterns across different time spans through multi-level projection, while also incorporating inter-stock relationships and market influences to improve stock movement prediction.

\section{Problem Definition}\label{chapter:p}
In this section, we will introduce some concepts in our proposed \name framework and formally define the problem of stock movement prediction. 
For certain concepts and phenomena, we provide examples to facilitate understanding.

\textbf{Definition 1. Stock Context.}
The set of all stocks is defined as $\mathcal{S}=\{s_1, s_2, ..., s_N\}\in\mathbb{R}^{N\times L\times F}$, where $s_i$ represents a specific stock, $N$ denotes the total number of stocks, $L$ denotes the length of the lookback window and $F$ is the number of features.
For any given stock $s_i$, its data on trading day $t$ is defined as $s_i^t\in\mathbb{R}^F$. Closing price $p_i^t$ is one of the features of $s_i^t$ and a one-day return ratio $r_i^t = \frac{p_i^t-p_i^{t-1}}{p_i^{t-1}}$.
On any given trading day $t$, there exists an optimal ranking of the stock scores $Y^t=\{y_1^t\geq y_2^t\geq...\geq y_N^t\}$. 
For any two stocks $s_i, s_j\in\mathcal{S}$, if $r_i^t\geq r_j^t$, there exists an overall order between the ranks $y_i^t\geq y_j^t$.
Such an ordering of stocks $\mathcal{S}$ on a trading day $t$ represents a ranking list, where stocks achieving higher ranking scores $Y$ are expected to achieve a higher
investment revenue (profit) on day $t$. 

\textbf{Definition 2. Industry Decay Matrix.}
Investors believe that firms with similar industry characteristics should earn similar returns on average~\cite{cisthpan}. The Industry Decay Matrix $D$ is designed to enhance the similarity characteristics within industries. Specifically, it assigns different decay coefficients to firms that belong to the same secondary industry $Se(\cdot)$, the same primary industry $Pr(\cdot)$, or different industries.

\begin{figure}[t]
    \centering
    \includegraphics[width=0.45\textwidth]{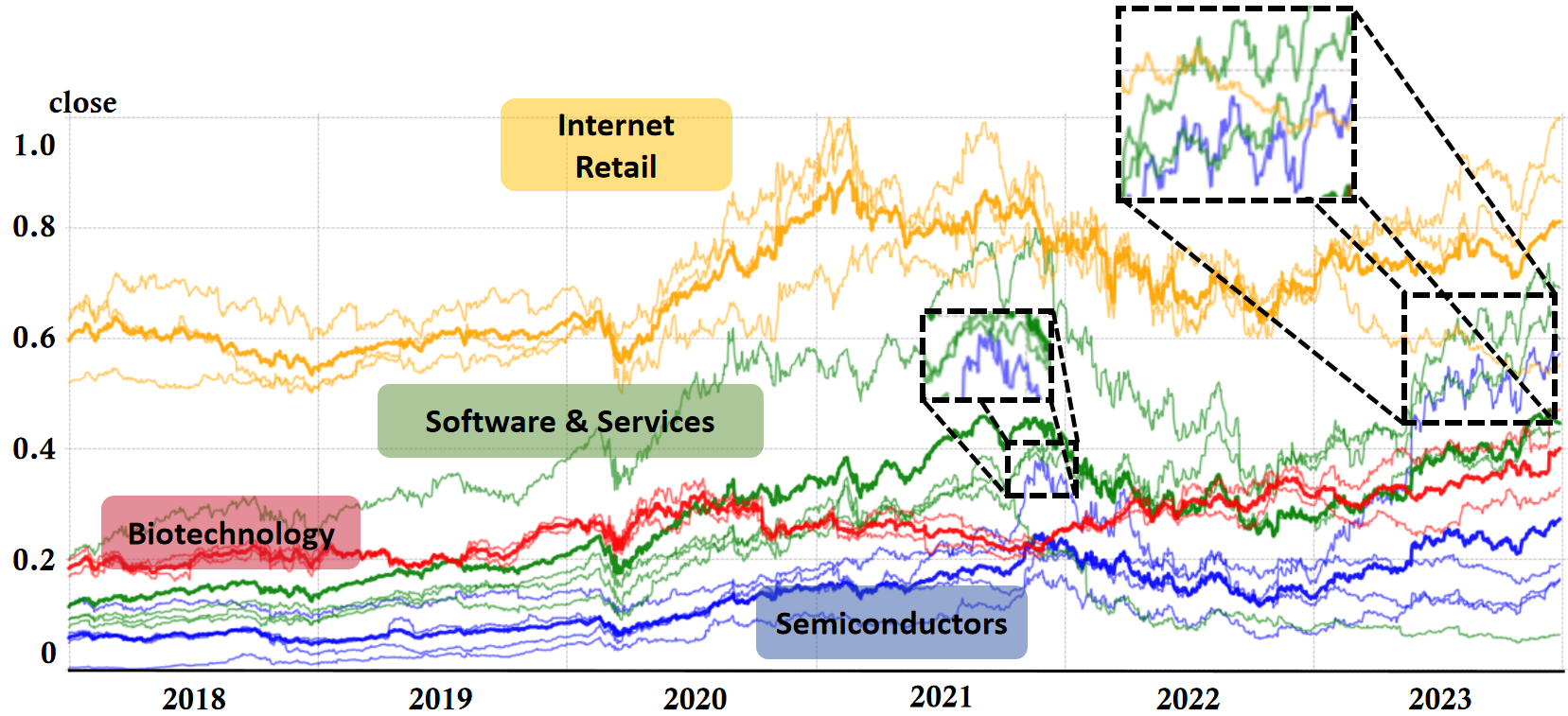}
    \caption{
         Stock movements among different primary and secondary industries.
    }
    \Description{..}
    \label{fig:industry}
\end{figure}

\textbf{Example 1.}
\cref{fig:industry} shows an example of fairly consistent movements among stocks within the same secondary industries. Moreover, within the same primary industries, such as Software \& Services and Semiconductors, there are also similar movements over certain periods, reflecting higher-level relationships.

\textbf{Definition 3. Dynamic Stock Correlation Graph.}
Given that stock relationships are subject to daily changes and shaped by market dynamics, we propose a Dynamic Stock Correlation Graph (DSCG) to capture and represent them. Let's define the DSCG as $\mathcal{G}=\{g^t\}_{t=1}^L$, where $g^t=\{\mathcal{V}, \mathcal{E}^t\}$ represents a specific stock correlation for one day. In graphs, the node set $\mathcal{V}$ denotes each stock and the edge set $\mathcal{E}$ represents stock correlations. 
Each node $v_i$ corresponds to a stock $s_i$.
Each edge $e_{ij}^t$ is assigned a weight $A^t[i,j]$, representing the relationship between $s_i$ and $s_j$ on trading day $t$, where $A^t$ is the adjacency matrix. The edges with the top $K$ similarities indicate the dominant interactive relationships on trading day $t$.

\textbf{Definition 4. Market Index.}
A market index $M$ is defined as a statistical measure that tracks the performance of a specific segment of the macroeconomic financial market.

\begin{figure*}[!t]
    \centering
    \includegraphics[width=0.9525\textwidth]{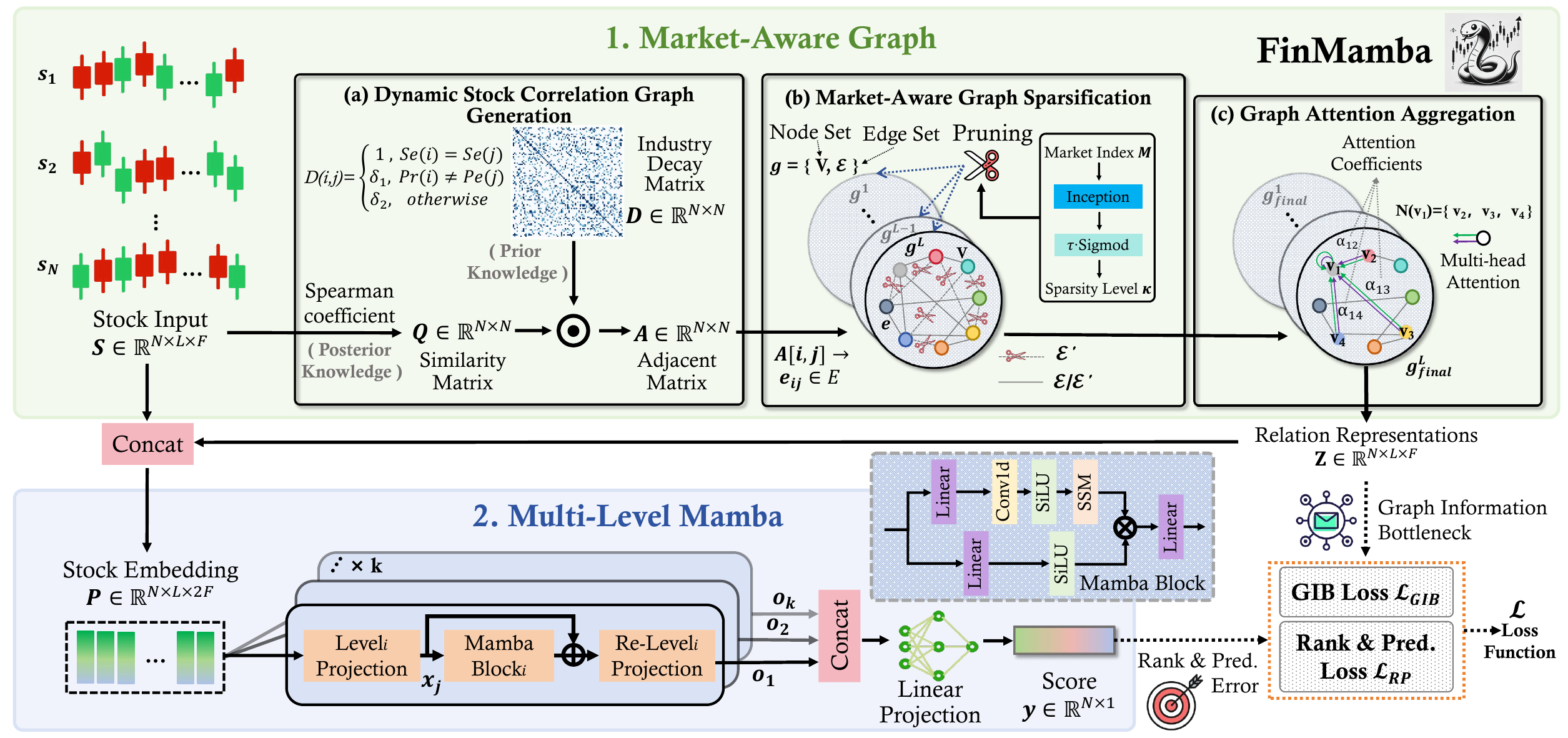}
    \caption{
    Overall structure of proposed \name. 
    1. Market-Aware Graph models both the prior long-term and the posterior short-term correlations between stocks and adaptively selects dominant relationships based on macroeconomic market index trends. 
    2. Multi-Level Mamba captures similar stock movement patterns across both coarse-grained and fine-grained levels.
    }
    \Description{..}
    \label{fig:pipline}
\end{figure*}

\textbf{Example 2.}
\cref{fig:corr&index} visualizes the correlation between the market index and individual stocks in the NASDAQ 100 from 2018 to 2023. A clear pattern emerges: \textbf{\textit{the stock correlation strengthens as the market index falls and weakens as the market index rises}}.
For instance, in March 2020, the COVID-19 outbreak in the U.S. revealed corporate debt risks and triggered a liquidity crisis, causing four circuit breakers in the stock market. 

We propose the following \textbf{\textit{hypothesis}} to explain this phenomenon:
(1) From the perspective of investors and risk concentration, when the market index declines, systemic risk rises and investor pessimism increases, leading to greater synchronization among stocks and higher correlation. Conversely, when the market performs well, optimism drives capital into diverse sectors, causing stock performance to diverge and reducing correlation.
(2) When the index rises, significant capital inflows into individual stocks can lower serial correlations, as diversified investment strategies spread buying interest, making price movements less synchronized.

\textbf{Problem 1. Stock Movement Prediction.}
Formally, given the stock-specific information (e.g., historical price, stock correlation) of $\mathcal{S}$, we aim to learn a ranking function that outputs a score $Y^{L+1}$ to rank each stock $s_i$ on the next day regarding expected profit.

\section{Method}

In the following sections, we will describe the architecture of the \name as illustrated in \cref{fig:pipline}, including the Market-Aware Graph and the Multi-Level Mamba.
Market-Aware Graph extracts the dependencies between stocks over a period of time under specific market conditions, including both the posterior short-term relationship between stock sequences and the prior long-term industry relationship, while Multi-Level Mamba effectively captures similar stock movement patterns at multiple levels.

\subsection{Market-Aware Graph}

Market-Aware Graph captures stock node representations from graphs generated on each trading day and consists of three key components: Dynamic Stock Correlation Graph Generation, Market-Aware Graph Sparsification, and Graph Attention Aggregation.

\paragraph{\textbf{Dynamic Stock Correlation Graph Generation}}
To model the comprehensive inter-stock correlations, represented by the adjacency matrix $A$, we approach from the perspective of the posterior sequence correlations $Q$ and the prior relationships $D$ of the primary and secondary industries.


In light of the fact that relationships between stocks are multifaceted and evolve on a daily basis~\cite{MDGNN}, it is evident that the methodologies employed in previous studies, which rely on prior static correlation graphs based on predefined concepts or rules~\cite{sgnn,sgnn1}, or those that generate posterior relationships directly from market history trends without incorporating additional domain-specific knowledge or news sources~\cite{thgnn, tcgpn, ECHOGL}, are inadequate to accurately capture the dynamic and real-time inter-stock correlations.

Given these considerations, we first calculate the posterior Spearman coefficient between the actual stocks $s_i,i\in\{1,2,...,N\}$ within a $L$-length look-back window to measure the interactions between the stocks, yielding the similarity matrix $Q\in\mathbb{R}^{N\times N}$. On trading day $t$, the similarity between $s_i$ and $s_j$ can be formalized with:

\begin{equation}
    Q_{ij}^{t}=1-\frac{6\sum_{c=t-L+1}^{t}(R(s_i^c)-R(s_j^c))^2}{L(L^2-1)}
\end{equation}
where $R(\cdot)$ is the rank function.

Subsequently, we propose the prior Industry Decay Matrix $D\in\mathbb{R}^{N\times N}$, which models the relationships from an industry perspective, thereby extending the analysis beyond the inherent correlations within financial sequences themselves. To model the variety of different types of relationships existing between companies (e.g., Franchisor and Franchisee, shareholders and invested companies), we introduce primary $Pr(\cdot)$ and secondary $Se(\cdot)$ industries, and define $D$ as follows:
\begin{equation}
D_{ij}=
\begin{cases}
    \ 1, \ \ \ \text{if}\ \ Se(i)=Se(j)\\
    \delta_1, \ \ \text{elif}\ \ Pr(i)=Pr(j)\\
    \delta_2, \ \ \text{otherwise}
\end{cases}
\end{equation}

Thus, $Q$ represents a posterior short-term relationship that changes daily, while $D$ is a constant prior long-term relationship. Ultimately, we obtain an adjacency matrix $A^t\in\mathbb{R}^{N\times N}$ by performing a dot product $A^t=Q^t\cdot D$, where $A^t[i,j]$ indicates an edge $e_{ij}^t\in\mathcal{E}^t$. 
Then a fully connected graph $g^t=\{\mathcal{V}, \mathcal{E}^t\}$, representing a comprehensive and refined correlations between stocks $S$, is generated.

\paragraph{\textbf{Market-Aware Graph Sparsification}}

In the long-term process of adapting to market changes, a fundamental observation of investors is that interactions between stocks evolve over time, dynamically activating and expiring. This means that certain relationships may effectively reflect similar stock movements in certain market conditions but not in others.
Traditional investors perform repeated statistical checks to identify relationships that are effective for portfolio selection, which is labour-intensive and challenging when combined with deep learning-based methods. To reduce manual effort, we propose the Market-Aware Graph Sparsification module, which incorporates market information to automatically select important inter-stock relationships.

Specifically, to ensure fairness without introducing additional information, for trading day $t$, the market index $M^t\in\mathbb{R}^{1\times L\times F}$ is represented by the mean value of the stocks included in the target set of length $L$ preceding that day, rather than utilising an actual market index. 
The process of calculating market-aware sparsity level is formalized as follows:
\begin{equation}
    \kappa^t=\tau\cdot \text{Sigmod}(\text{Inception}(M^t))
\end{equation}
where $\tau$ is a constant used to adjust the sparsity level. 
First, the 2D tensor $M$ is processed by a parameter-efficient inception block~\cite{inception}, designated as $Inception(\cdot)$, which incorporates multi-scale 2D kernels and is widely recognized as a prominent vision backbone. 
Subsequently, the output of $Inception(\cdot)$ passes through the $sigmoid$ activation function to obtain the pruning sparsity level $\kappa^t$. 

Finally, adaptive pruning is performed on the already generated fully connected graph $g^t$ to obtain the dominant $K$ correlations between stocks as perceived by the market over a certain period. The process of sparsification is formalized as follows:
\begin{equation}
    g_{final}^{t}=\{V,\mathcal{E}^t\backslash\mathcal{E}^{t'}\},\
    \mathcal{E}^{t'} = \text{topK}(-A^t[i,j],\lceil \kappa^t \times |\mathcal{E}^{t}| \rceil)
\end{equation}
where $g_{final}^t$ only modifies the edge set $\mathcal{E}^t$ without altering the node set $\mathcal{V}$, and $\mathcal{E}^{t'}$ denotes the removed edges. The retained edges $e_{ij}^t\in\mathcal{E}^t\backslash\mathcal{E}^{t'}$ correspond to the edges with the top $K$ largest weights $A^t[i,j]$, where $K=|\kappa^t\cdot N^2|$.

\paragraph{\textbf{Graph Attention Aggregation}}
After generating a set of DSCG $\mathcal{G}$, we leverage a multi-head attention mechanism to aggregate the messages from neighboring nodes within graph structures. 
Concretely, for each stock $s_i$ (node $v_i$) on the trading day $t$, we compute the attention coefficient $\alpha_{ij}^t$ of its neighboring stock $s_j$ (node $v_j$), denoting the importance of the edge $e_{ij}^t$ in the pruned graph $g_{final}^t$. It can formalized with:
\begin{equation}
    \alpha_{v_i, v_j}^{k} = \frac{\text{exp}(\text{LeakyReLU}(a_k^T[Wh_i^t||Wh_j^t]))}{\sum_{s_u^t\in\mathcal{N}(v_i)}\text{exp}(\text{LeakyReLU}(a_k^T[Wh_i^t||Wh_u^t]))}
\end{equation}
where $h_i^t, h_j^t\in\mathbb{R}^{L\times F}$ is the feature representation of nodes $v_i$ and $v_j$, $W\in\mathbb{R}^F$ is the learnable weight matrix, $a_k^T\in\mathbb{R}^{2F}$ is the weight vector of relation $e_{ij}^t$ in $k$-th head and function $\mathcal{N}(v_i)$ denotes the set of neighbors of node $v_i$.

Then we aggregate the features of neighboring nodes of node $v_i$ using attention coefficients to generate a neighboring representation $z_i^t\in\mathbb{R}^{1\times L\times F}$:


\begin{equation}
z_i^t=\sigma\bigg(\bigg|\bigg|_{k=1}^{K}\sum_{v_j\in\mathcal{N}(v_i)}\alpha_{v_i, v_j}^{k}h_j^t\bigg)
\end{equation}

where $\sigma$ denotes a non-linear activation function GELU and $\big|\big|_{k=1}^{K}$ denotes the concatenation of the outputs from all $K$ heads.

Once we have the original representations $s_i^t$ and the neighboring representation $z_i^t$, we concatenate them and get the stock embedding $P=\text{Stack}(s_i^t||z_i^t)|_{i=1}^{N}\ \in\mathbb{R}^{N\times L\times 2F}$.

\subsection{Multi-Level Mamba}

Compared to other financial time series forecasting methods (e.g., RNN, CNN, Transformer), Mamba introduces a input-dependent selection mechanism that balances short-term and long-term dependencies, which is suitable for modeling stock data where similar stock patterns occur over different time spans.
In earlier modules, we have already generated sequences that capture both short-term and long-term relationships, providing a strong foundation for this adaptive mechanism.
However, stock data exhibit different temporal variations across various scales. Intuitively, in a stock price series, daily fluctuations are influenced by monthly economic trends, which in turn are influenced by annual market cycles. To better model the diverse patterns within the non-stationary dynamics of stocks, we designed a Multi-Level Mamba.

Technically, given the input $P\in\mathbb{R}^{N\times L\times 2F}$, we project it onto $k$ different levels through $k$ distinct linear mappings, each designed to capture features at varying levels of granularity.
These mappings, parameterized by unique transformation matrices, allow the model to extract multi-level pattern-oriented information by emphasizing either fine-grained local details or broader global patterns, enabling a richer representation.
For each level $i\in\{1,2,...,k\}$, it is fed into Mamba block, where the continuous state space mechanism produces a response $o^t_i\in\mathbb{R}^{N\times D}$ based on the observation of hidden state $h^t_i\in\mathbb{R}^{N\times D}$ and the input $x_i^t=\text{Linear}_i(P)^t\in\mathbb{R}^{N\times 2F}$, where $D$ is the output dimension. It can be formulated as:
\begin{equation}
\begin{aligned}
    h^t_i &= \bar{A}^t_i h^{t-1}_{i}+\bar{B}^t_i x_i^t\\
    o^t_i &= C^t_i h^t_i
\end{aligned}
\end{equation}
where $\bar{A}^i_t$, $\bar{B}^t_i$ and $C^t_i$ are the parameters of Mamba block~\cite{mamba}.

Afterward, we apply another linear mapping to restore the original level and concatenate them. Finally, a linear layer is used to obtain the predicted score $y^t\in\mathbb{R}^{N\times 1}$ of next trading day and $Y$ is formed by concatenating $y^t$ along the time dimension.
\begin{equation}
    y^t = \text{Linear}(||_{i=1}^{k}[\text{Re-Level-Proj.}(o^t_i+x_i^t)])
\end{equation}

Empowered by the multi-level state space mechanism, \name learns similar stock patterns over random intervals at different levels, enabling it to integrate complementary forecasting capabilities from mixed multi-level series.

\renewcommand{\arraystretch}{0.95}
\begin{table*}[!ht]
\setlength{\tabcolsep}{2pt}
\caption{Performance evaluation of compared models for financial time series forecasting in CSI 300, CSI 500, S\&P 500 and NASDAQ 100 datasets. The best and second-best results are in \best{bold} and \second{underlined}, respectively.}
\resizebox{\textwidth}{!}
{
\begin{tabular}{c|ccccc|ccccc|ccccc|ccccc}
\toprule
 \multicolumn{1}{c}{\multirow{2}{*}{Models}} & \multicolumn{5}{c}{CSI 300} & \multicolumn{5}{c}{CSI 500} & \multicolumn{5}{c}{S\&P 500}& \multicolumn{5}{c}{NASDAQ 100} \\
      
\cmidrule(lr){2-6} \cmidrule(lr){7-11} \cmidrule(lr){12-16} \cmidrule(lr){17-21}

\multicolumn{1}{c}{} & ARR$\uparrow$ & AVol$\downarrow$ & MDD$\downarrow$ & ASR$\uparrow$ & IR$\uparrow$ & ARR$\uparrow$ & AVol$\downarrow$ & MDD$\downarrow$ & ASR$\uparrow$ & IR$\uparrow$ & ARR$\uparrow$ & AVol$\downarrow$ & MDD$\downarrow$ & ASR$\uparrow$ & IR$\uparrow$ & ARR$\uparrow$ & AVol$\downarrow$ & MDD$\downarrow$ & ASR$\uparrow$ & IR$\uparrow$ \\
\midrule
BLSW~\citeyearpar{blsw} & -0.076 & \best{0.113} & -0.231 & -0.670 & \second{0.311} & 0.110 & 0.227 & -0.155 & 0.485 & 0.446 & 0.199 & 0.318 & -0.223 & 0.626 & 0.774 & 0.368 & 0.339 & -0.222 & 1.086 & 1.194 \\
CSM~\citeyearpar{csm} & -0.185 & 0.204 & -0.293 & -0.907 & -0.935 & 0.015 & 0.229 & -0.179 & 0.066 & 0.001 & 0.099 & 0.250 & -0.139 & 0.396 & 0.584 & 0.116 & 0.242 & -0.145 & 0.479 & 0.603 \\
\midrule
AlphaStock~\citeyearpar{alphastock} & -0.164 & 0.153 & -0.245 & -1.072 & -1.098 & -0.017 & 0.148 & -0.166 & -0.115 & -0.043 & 0.122 & 0.140 & -0.126 & 0.871 & 0.892 & 0.372 & 0.178 & -0.134 & 1.781 & 1.869 \\
DeepPocket~\citeyearpar{deeppocket} & -0.036 & 0.135 & -0.175 & -0.270 & -0.258 & 0.006 & \best{0.127} & -0.148 & 0.050 & 0.115 & 0.165 & \second{0.142} & -0.126 & 1.165 & 1.045 & 0.346 & \best{0.157} & -0.116 & 2.197 & 1.882 \\
\midrule
Transformer~\citeyearpar{attention} & -0.240 & 0.156 & -0.281 & -1.543 & -1.695 & 0.154 & 0.156 & -0.135 & 0.986 & 0.867 & 0.135 & 0.159 & -0.140 & 0.852 & 0.908 & 0.268 & 0.175 & -0.131 & 1.531 & 1.441 \\
Mamba~\citeyearpar{mamba} & -0.141 & 0.155 & -0.290 & -0.910 & -1.240 & 0.076 & 0.157 & -0.180 & 0.484 & 0.337 & 0.141 & 0.170 & -0.156 & 0.829 & 0.892 & 0.281 & 0.170 & -0.143 & 1.651 & 1.504 \\
FactorVAE~\citeyearpar{FactorVAE} & -0.048 & \second{0.134} & -0.175 & -0.335 & -0.348 & 0.006 & \best{0.127} & -0.147 & 0.047 & 0.112 & 0.160 & \second{0.142} & -0.132 & 1.128 & 1.013 & 0.356 & \second{0.159} & -0.119 & 2.234 & 1.907 \\
THGNN~\citeyearpar{thgnn} & -0.015 & 0.172 & -0.152 & \second{-0.088} & -0.003 & 0.048 & \second{0.128} & -0.141 & 0.375 & 0.432 & 0.271 & \best{0.141} & -0.094 & 1.921 & 1.778 & 0.644 & 0.204 & -0.146 & 3.147 & 2.543 \\ 
MambaStock~\citeyearpar{mambastock} & -0.132 & 0.158 & -0.227 & -0.836 & -0.857 & 0.106 & 0.154 & -0.158 & 0.690 & 0.339 & 0.145 & 0.158 & -0.146 & 0.916 & 0.932 & 0.280 & 0.166 & -0.140 & 1.684 & 1.489 \\
CL~\citeyearpar{CL} & -0.035 & 0.148 & -0.183 & -0.241 & -0.193 & 0.051 & 0.146 & -0.128 & 0.351 & 0.390 & 0.308 & 0.189 & -0.171 & 1.629 & 1.451 & 0.351 & 0.172 & -0.122 & 2.041 & 1.821 \\
MASTER~\citeyearpar{master} & \second{0.102} & 0.151 & \second{-0.126} & 0.681 & 0.726 & 0.128 & 0.130 & \best{-0.098} & 0.989 & 0.997 & \second{0.335} & 0.171 & -0.134 & \second{1.958} & \second{1.895} & \second{0.654} & 0.188 & -0.102 & \second{3.479} & 2.683 \\ 
CI-STHPAN~\citeyearpar{cisthpan} & -0.078 & 0.167 & -0.144 & -0.466 & -0.355 & 0.021 & 0.151 & -0.129 & 0.136 & 0.211 & 0.123 & 0.233 & -0.254 & 0.527 & 0.632 & 0.454 & 0.208 & -0.124 & 2.178 & 1.855 \\ 
VGNN~\citeyearpar{VGNN} & -0.037 & 0.163 & -0.197 & -0.227 & -0.201 & 0.111 & 0.166 & -0.175 & 0.668 & 0.564 & 0.299 & 0.202 & -0.169 & 1.473 & 1.406 & 0.616 & 0.181 & \second{-0.099} & 3.405 & \second{2.798} \\ 
\midrule
PatchTST~\citeyearpar{patchtst} & -0.224 & 0.158 & -0.279 & -1.415 & -1.563 & 0.118 & 0.152 & -0.127 & 0.776 & 0.735 & 0.146 & 0.167 & -0.140 & 0.877 & 0.949 & 0.239 & 0.185 & -0.138 & 1.296 & 1.233 \\
iTransformer~\citeyearpar{itransformer} & -0.115 & 0.145 & -0.190 & -0.793 & -0.775 & \second{0.214} & 0.168 & -0.164 & \second{1.276} & \second{1.173} & 0.159 & 0.170 & -0.139 & 0.941 & 0.955 & 0.188 & 0.196 & -0.202 & 0.963 & 0.937 \\
TimeMixer~\citeyearpar{timemixer} & -0.156 & 0.159 & -0.232 & -0.983 & -1.028 & 0.078 & 0.153 & -0.114 & 0.511 & 0.385 & 0.254 & 0.162 & -0.131 & 1.568 & 1.448 & 0.264 & 0.188 & -0.131 & 1.401 & 1.346 \\
Crossformer~\citeyearpar{crossformer} & -0.071 & 0.162 & -0.437 & -0.383 & -0.039 & 0.163 & 0.217 & -0.238 & 0.686 & 0.650 & 0.284 & 0.159 & \second{-0.114} & 1.786 & 1.646 & 0.363 & 0.181 & -0.167 & 2.010 & 1.860 \\
\midrule
\name (Ours) & \best{0.106} & 0.163 & \best{-0.102} & \best{0.653} & \best{0.705} & \best{0.227} & 0.163 & \second{-0.106} & \best{1.389} & \best{1.339} & \best{0.341} & 0.164 & \best{-0.086} & \best{2.063} & \best{1.937} & \best{0.685} & 0.190 & \best{-0.097} & \best{3.601} & \best{2.851} \\
\bottomrule
\end{tabular}
}
\label{tab:mainexp}
\end{table*}
\renewcommand{\arraystretch}{1.0}

\subsection{Optimization Objectives}
In stock movement prediction, the learning goal of \name is to estimate the predicted $y^t$ denoting the positive return of all $N$ stocks on trading day t. For overall optimization, we combine point-wise regression loss and pair-wise ranking loss as:
\begin{equation}
    \mathcal{L}_{RP}=\frac{1}{L}\sum_{t=1}^{L}(\sum_{i=1}^{N}||y_i^t-r_i^t||^2+\eta\sum_{i=1}^{N}\sum_{j=1}^{N}max(0,-(y_i^t-y_j^t)(r_i^t-r_j^t)))
\end{equation}

Moreover, to guarantee that the learned stock embeddings efficiently capture the correlation within the dynamic graph structure and address the graph information bottleneck~\cite{gib} issue by minimizing uncertainty in the graph, we incorporate GIB loss $\mathcal{L}_{GIB}$ to minimize mutual information between the aggregation embedding and the original input, ensuring key information is retained while reducing irrelevant details.

\begin{equation}
    \mathcal{L}_{GIB}=I(Z;X)=\frac{1}{L}\sum_{t=1}^{L} \bigg(\sum_{i=1}^{N}\frac{||\text{mean}(z_i^t)-\text{mean}(s_i^t)||^2}{\text{var}(z_i^t)+\text{var}(s_i^t)}\bigg)
\end{equation}

Combining these two loss, we reach the complete end-to-end loss function with a weighting coefficient $\lambda$:
\begin{equation}
    \mathcal{L}=\mathcal{L}_{RP} + \lambda\mathcal{L}_{GIB}
\end{equation}

\section{Expeiments}

\subsection{Experiment Setup}

\paragraph{\textbf{Datasets.}}
We conduct thorough experiments in both the U.S. and Chinese stock markets, selecting entities from the 
S\&P 500\footnote{\url{https://hk.finance.yahoo.com/quote/\%5EGSPC/history/}}, 
NASDAQ 100\footnote{\url{https://hk.finance.yahoo.com/quote/\%5EIXIC/history}}, 
CSI 300\footnote{\url{https://cn.investing.com/indices/csi300}}, and 
CSI 500\footnote{\url{https://cn.investing.com/indices/china-securities-500}}. Our datasets comprise historical day-level market information, such as close, open, high, low, turnover and volume, from 2018 to 2023. See \cref{app:data} for a more detailed description of the datasets.


\paragraph{\textbf{Baseline Models.}}
We compare \name with other competitive models from different categories as follows: 
(1) Quantitative Investment Methods:
a. Classic strategies: Buying-Loser-Selling-Winner (BLSW)~\cite{blsw} and Cross-Sectional Mean reversion (CSM)~\cite{csm}; 
b. Deep Reinforcement Learning methods: AlphaStock~\cite{alphastock}, DeepPocket~\cite{deeppocket};
c. Deep Learning methods: Transformer~\cite{attention}, Mamba \cite{mamba}, FactorVAE~\cite{FactorVAE}, THGNN~\cite{thgnn}, MambaStock~\cite{mambastock}, CL~\cite{CL}, MASTER~\cite{master}, CI-STHPAN~\cite{cisthpan} and VGNN~\cite{VGNN};
(2) General Time Series Forecasting methods: PatchTST \cite{patchtst}, iTransformer~\cite{itransformer}, TimeMixer \cite{timemixer}, and Crossformer~\cite{crossformer}.

\paragraph{\textbf{Metric.}}
We employ five widely used metrics to evaluate the overall performance of each method: Annual Return Ratio (ARR), Annual Volatility (AVol), Maximum Draw Down (MDD), Annual Sharpe Ratio (ASR), and Information Ratio (IR). Lower absolute values of AVol and MDD, along with higher ARR, ASR, and IR, indicate better performance. See \cref{app:metric} for detailed descriptions.

\paragraph{\textbf{Implementation Details.}}
Our experiment is trained on the NVIDIA V100 GPU, and all models are built using PyTorch \cite{pytorch}.
The training and validation sets are kept consistent for all models. The number of GNN layers is 2, the number of levels is 2, and the window size is 20. See \cref{app:imp} for detailed settings.

\renewcommand{\arraystretch}{0.85}
\begin{table*}[t]
\caption{Component ablation of \name in CSI 500 and NASDAQ 100 datasets.}
\resizebox{0.95\textwidth}{!}
{
\begin{tabular}{c|cccccc|cccccc}
\toprule
 \multicolumn{1}{c}{\multirow{2}{*}{Models}} & \multicolumn{6}{c}{CSI 500} & \multicolumn{6}{c}{NASDAQ 100} \\
      
\cmidrule(lr){2-7} \cmidrule(lr){8-13}

\multicolumn{1}{c}{} & ARR$\uparrow$ & AVol$\downarrow$ & MDD$\downarrow$ & ASR$\uparrow$ & CR$\uparrow$ & \multicolumn{1}{c}{IR$\uparrow$} & ARR$\uparrow$ & AVol$\downarrow$ & MDD$\downarrow$ & ASR$\uparrow$ & CR$\uparrow$ & IR$\uparrow$  \\
\midrule
w/o industry decay matrix & 0.146 & 0.165 & -0.111 & 0.887 & 1.312 & 0.914 & 0.677 & 0.190 & -0.101 & 3.559 & 6.717 & 2.825 \\
w/o Market-Aware Sparsification & 0.064 & 0.162 & -0.186 & 0.391 & 0.341 & 0.462 & 0.399 & 0.184 & -0.112 & 2.166 & 3.543 & 1.923 \\
w/o Dynamic Stock Correlation & 0.128 & 0.168 & -0.113 & 0.757 & 1.125 & 0.799 & 0.384 & 0.188 & -0.115 & 2.043 & 3.351 & 1.831 \\
single level Mamba & 0.129 & 0.163 & -0.110 & 0.789 & 1.172 & 0.827 & 0.432 & 0.184 & -0.115 & 2.347 & 3.754 & 2.052 \\
w/o Mamba & 0.046 & \best{0.148} & -0.130 & 0.310 & 0.353 & 0.378 & 0.312 & \best{0.179} & -0.149 & 1.743 & 2.098 & 1.631 \\
FinTransformer & 0.194 & 0.168 & -0.144 & 1.152 & 1.352 & 1.142 & 0.671 & 0.183 & -0.117 & \best{3.676} & 5.726 & 2.717 \\
\midrule
\name & \best{0.227} & 0.163 & \best{-0.106} & \best{1.389} & \best{2.135} & \best{1.339} & \best{0.685} & 0.190 & \best{-0.097} & 3.601 & \best{7.085} & \best{2.851} \\
\bottomrule
\end{tabular}
}
\label{tab:ablation}
\end{table*}
\renewcommand{\arraystretch}{1.0}

\begin{figure*}[!htb]
    \centering
    \begin{subfigure}[!b]{0.73\textwidth}
        \centering
        \includegraphics[width=\textwidth]{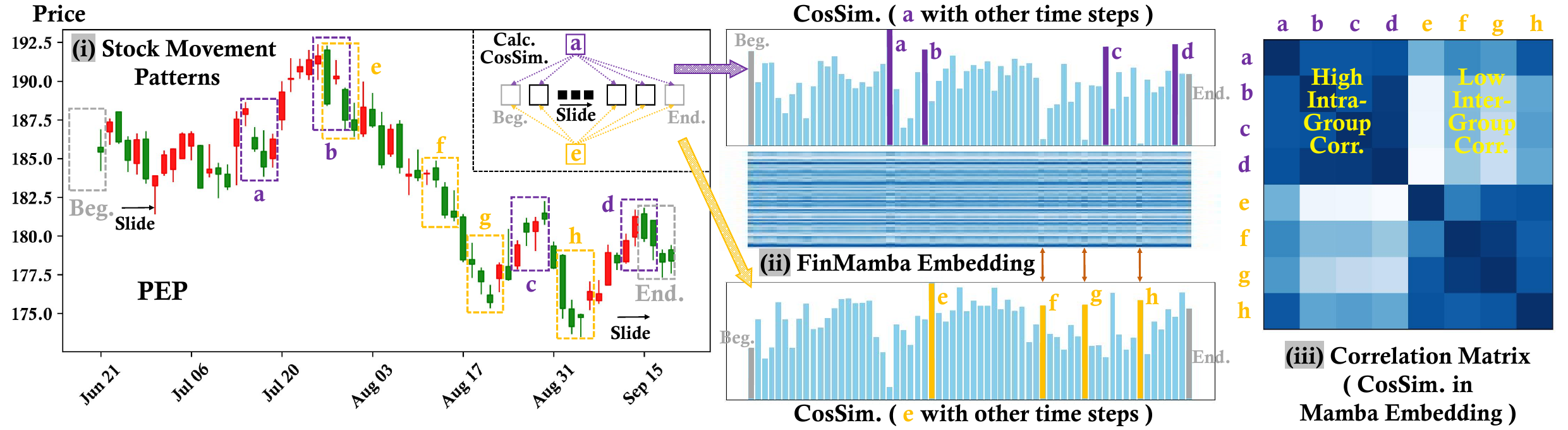} 
        \caption{}
        \label{fig:subfig1}
    \end{subfigure}
    \hfill
    \begin{subfigure}[!b]{0.26\textwidth}
        \centering
        \includegraphics[width=\textwidth]{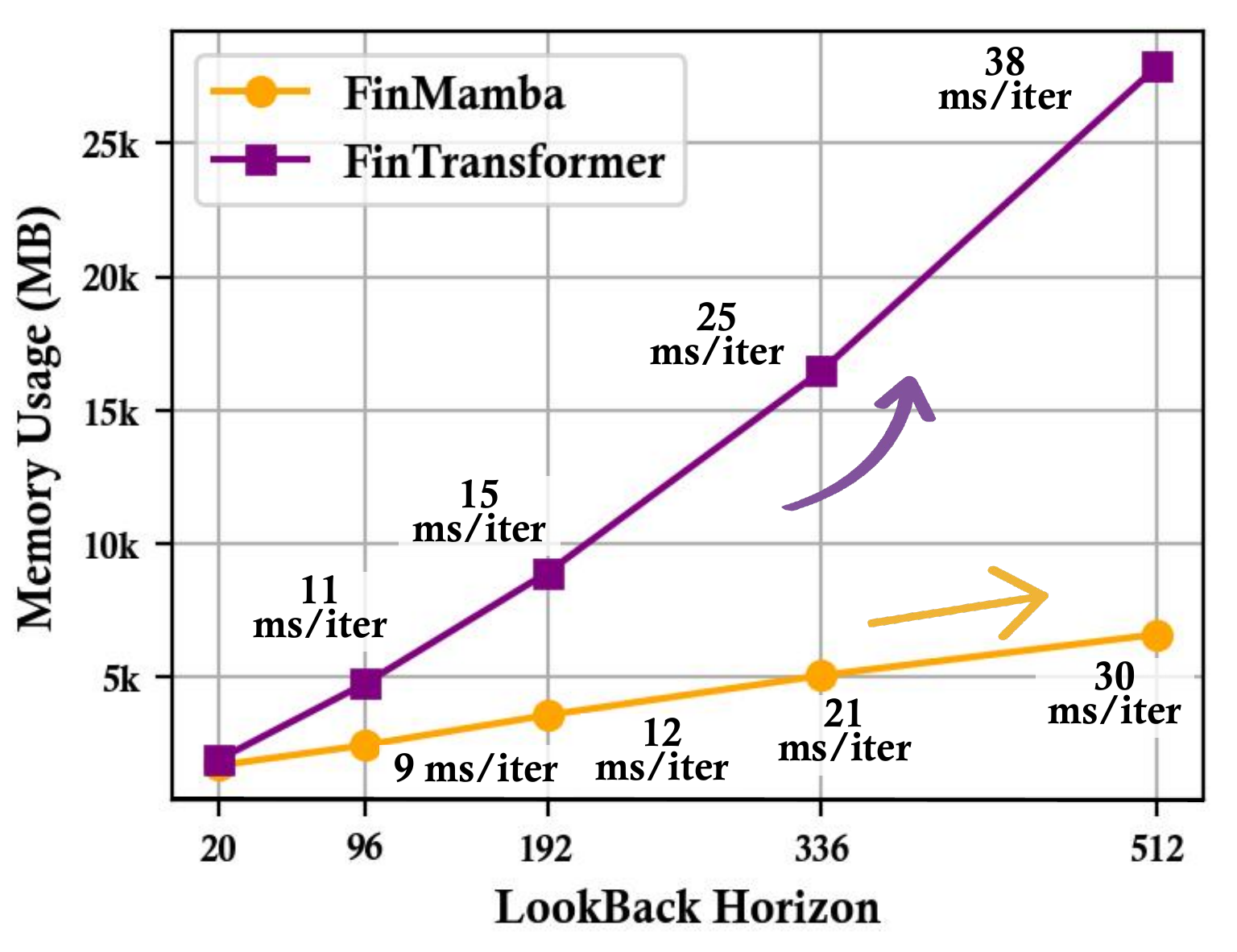} 
        \caption{}
        \label{fig:subfig2}
    \end{subfigure}
    \caption{(a) \textcolor{mypurple}{a} to \textcolor{mypurple}{d} and \textcolor{myorange}{e} to \textcolor{myorange}{h} represent two groups of similar stock movement patterns. \name effectively captures strong intra-group correlations and exhibits low inter-group correlations. (b) Our proposed \name demonstrates superior efficiency and effectiveness, achieving lower inference time (ms/iter) and reduced memory usage (MB) with longer lookback horizons.}
    \label{fig:mamba_trans}
    \Description{..}
\end{figure*}

\paragraph{\textbf{Trading Protocols.}}
Following THGNN~\cite{thgnn}, we use the daily buy-hold-sell trading strategy to evaluate the performance of stock movement prediction methods in terms of returns. 
During the test period, the trading process for each day is simulated as follows: First, at the end of the trading on day $t$, the traders use \name to generate prediction scores and rank expected returns for all stocks. Then, at the opening of $t+1$, the traders sell the stocks purchased on day $t$ and buy those with higher predicted returns, focusing on the top $k$ ranked stocks. If a stock continues to be ranked with the highest predicted returns, it remains in the portfolio. Notably, the experiment does not take transaction costs into account.

\subsection{Experiment Results}



The overall performance is reported in \cref{tab:mainexp}. Our proposed \name outperforms all other methods on most metrics. Based on these experimental findings, we draw the following conclusions:

\paragraph{\textbf{TSF methods.}} 
Deep learning-based general time series forecasting (TSF) methods, such as Crossformer~\cite{crossformer} and iTransformer~\cite{itransformer}, do not exploit the graph relationships between stocks, making it difficult for them to achieve state-of-the-art performance.
Moreover, such TSF methods rely primarily on historical data to predict future trends, lacking real-time decision-making and dynamic feedback mechanisms. Given the complexity of stock markets, relying solely on predictions is often insufficient to mitigate risk.


\paragraph{\textbf{Graph-based quantitative investment methods.}} 
Graph-based stock prediction models, such as THGNN~\cite{thgnn}, VGNN~\cite{VGNN} and CI-STHPAN~\cite{cisthpan} perform exceptionally well, demonstrating that incorporating relationships between stocks can significantly improve predictive performance. By leveraging graph structures to simulate synergies, these models capture the complex interactions and dependencies that exist between different stocks.

\paragraph{\textbf{Marker-Aware quantitative investment methods.}} 
The proposed \name refines the inter-stock dependencies in the global set of stocks based on market feedback, while MASTER~\cite{master} employs a gating mechanism to select effective factors for individual stocks, achieving \best{optimal} and \second{suboptimal} results respectively. This demonstrates the powerful ability of market information to optimize portfolio strategies. It is important to note that \name performs well in ARR, ASR, CR, and IR across all four datasets, highlighting its superiority in stock movement prediction. \name also shows strong performance in MDD, confirming that market-aware graph effectively captures complex, non-linear market behaviors by incorporating interactions with the market environment.

\subsection{Ablation Study}

To evaluate the effectiveness of each module in \name, we perform extensive ablation experiments on the CSI 500 and NASDAQ 100 datasets, with results shown in \cref{tab:ablation}. 
Excluding the industry decay matrix $D$ (w/o industry decay matrix) reduces performance, confirming the importance of our primary and secondary industry-related graph construction. 
Omitting Market-Aware Sparsification (w/o Market-Aware Sparsification) leads to a significant drop in performance, highlighting the module's role in leveraging the relationship between the macroeconomic market index and whole stocks set and eliminating unnecessary noise introduced by irrelevant edges. 
Furthermore, ignoring stock relationships and treating Mamba purely as a time-series prediction task (w/o Dynamic Stock Correlation) or ignoring time-series information and using GNN purely for aggregating stock relationships (w/o Mamba) both result in performance degradation.

For a detailed comparison between Transformer~\cite{attention} and Mamba \cite{mamba},
we first replace Mamba in \name with Transformer (\textit{FinTransformer}) for experiments, resulting in a slight performance decline.
Subsequently, we visualize the model embeddings of the PEP stock from 23/06/21 to 23/09/20 in \cref{fig:subfig1} (i).
In the daily candlestick chart, we identify two groups of similar stock movement patterns: \textcolor{mypurple}{a} to \textcolor{mypurple}{d} and \textcolor{myorange}{e} to \textcolor{myorange}{h}. 
In the \name embedding, we calculate the cosine similarity between pattern \textcolor{mypurple}{a} and all other patterns, and also for pattern \textcolor{myorange}{e}, as shown in \cref{fig:subfig1} (ii). The correlation matrix of the eight patterns is presented in \cref{fig:subfig1} (iii).
We observe that \name effectively captures strong intra-group correlations, modeling well the temporal dependencies within similar patterns, and exhibits low inter-group correlations, reducing the impact of noise.
This suggests that \name can selectively retain similar inputs across different time intervals, making it highly effective for modeling stock data where recurring patterns may appear over varying time spans.
Lastly, we evaluate the model complexity in terms of memory usage (MB) and inference time (ms/iter) for longer lookback horizons, as depicted in \cref{fig:subfig2}. The lightweight and efficient design of \name, with its linear complexity, meets the timeliness requirement of algorithmic trading~\cite{hulandscape}.

\begin{figure}[!t]
    \centering
    \includegraphics[width=0.48\textwidth]{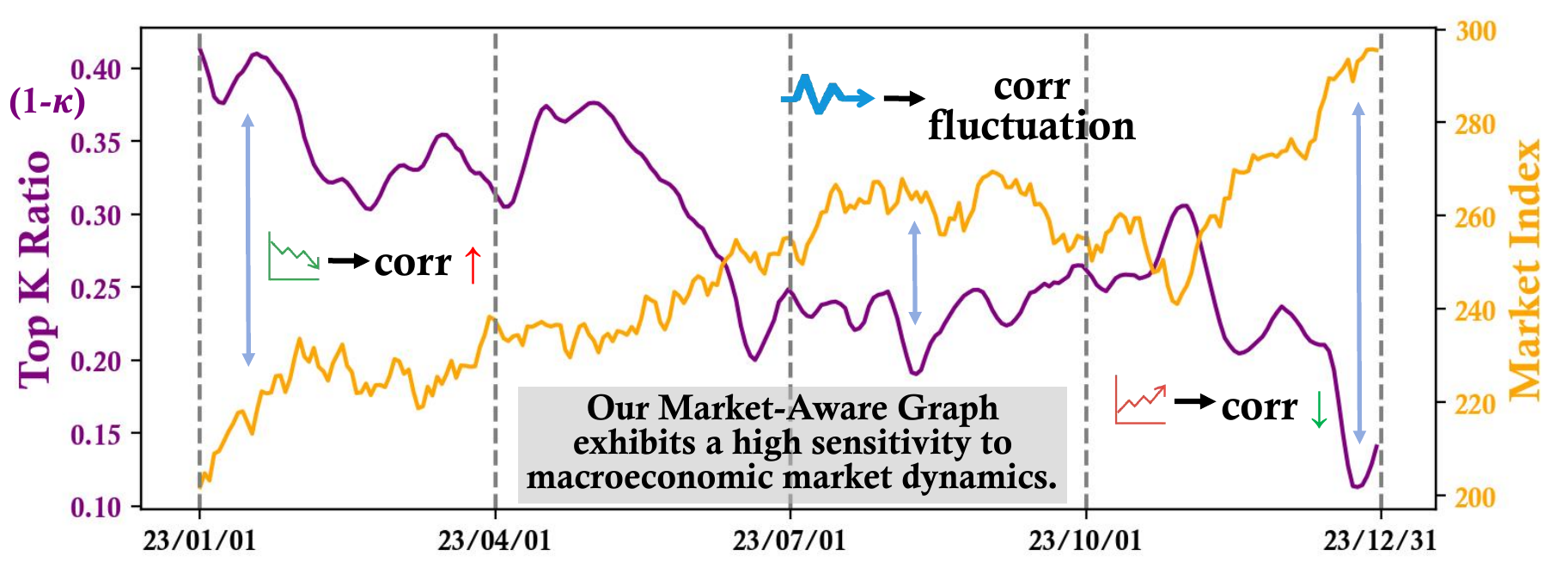}
    \caption{
         The retaining topK ratio with market index.
    }
    \Description{..}
    \label{fig:topk_ratio}
\end{figure}

\begin{figure}[!t]
    \centering
    \includegraphics[width=0.48\textwidth]{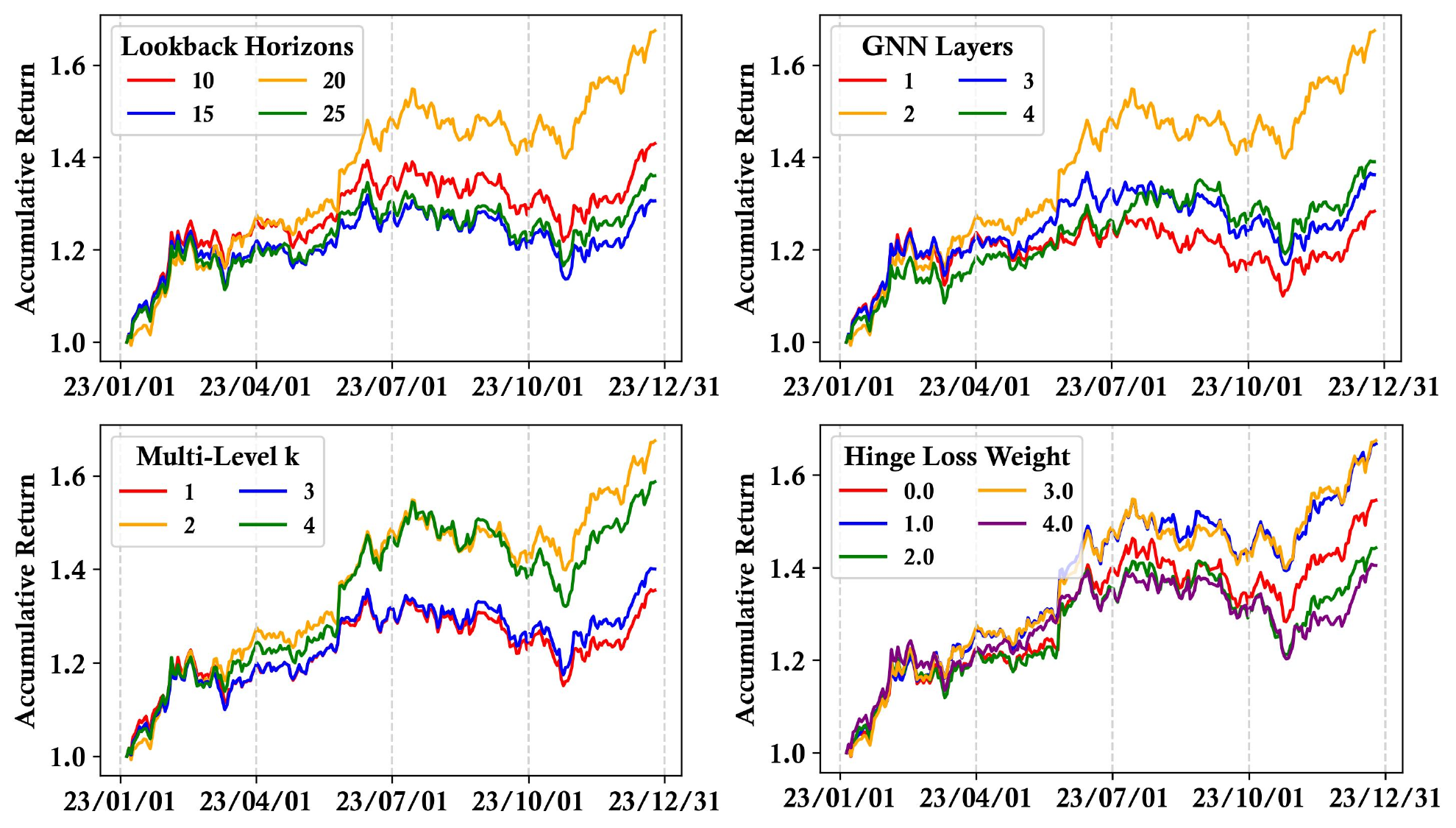}
    \caption{
         Portfolio performance of \name in terms of the length of lookback horizon $L$, the number of GNN layers, the weight of hinge loss $\eta$, and the number of levels $k$ in MLM.
    }
    \Description{..}
    \label{fig:para}
\end{figure}

\subsection{Sparsity Level $\kappa$ in Market-Aware Graph}

As shown in the \cref{fig:topk_ratio}, our Market-Aware Graph Sparsification is highly effective. When the market index is high, the relationships between stocks become weaker and it tends to retain fewer edges. Conversely, when the market index is low, the interactions between stocks become stronger and it tends to retain more edges.

\subsection{Parameter Sensitivity}
\cref{fig:para} shows the effect of four hyperparameters on the NASDAQ 100 dataset. It appears that increasing lookback horizons enhances performance, but only up to a certain length, beyond which the additional input introduces more noisy signals, diminishing the effectiveness of information capture. A similar pattern occurs when increasing the number of GNN layers and the number of levels in MLM. In addition, since hinge loss guides the model more effectively in learning ranking information, it improves ranking performance but is constrained by MSE loss, which limits prediction accuracy.

\subsection{Case Study}

\begin{figure}[!t]
    \centering
    \includegraphics[width=0.45\textwidth]{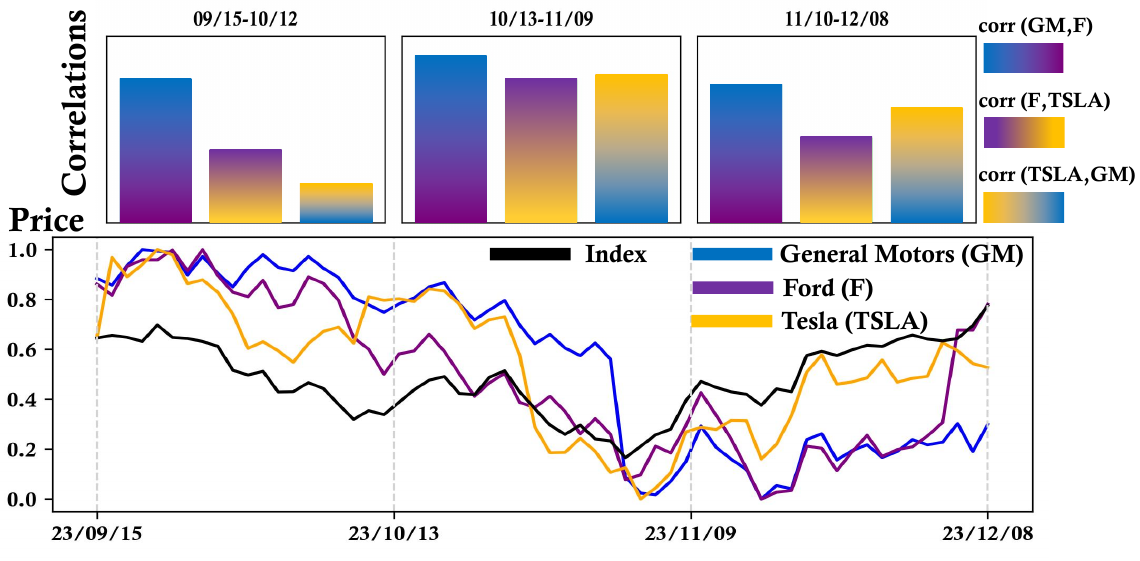}
    \caption{
         Case analysis of multifaceted inter-stock correlations. The upper half represents the correlation between the three stocks over different periods, while the lower half represents the stock price movements of the three stocks.
    }
    \Description{..}
    \label{fig:case}
\end{figure}

On 15 September 2023, the United Auto Workers (UAW) initiated its first-ever simultaneous strike against the ``Detroit Big Three'' automakers: General Motors, Ford, and Stellantis. Lasting a month and a half, the strike is expected to hinder these companies' ability to compete with non-union automakers like Tesla and foreign brands. This case study examines the stock market's response, focusing on the relationship between GM (General Motors), F (Ford) and TSLA (Tesla).
As in \cref{fig:case}, the correlation between GM and F remained strong throughout all three periods, reflecting their close ties as traditional auto giants. In contrast, TSLA's correlation with GM and F was weaker in the first period, perhaps reflecting its distinct market position and growth prospects. However, in the subsequent periods, as the strike continued and competition intensified, the relationship between TSLA and the other two stocks notably strengthened. This suggests that \name more accurately captures the evolving dynamics between the stocks, reflecting their interdependencies.

\vspace{-0.28em}

\section{Conclusion}

In this paper, we introduce a novel framework, \name, which integrates a Market-Aware Graph and Multi-Level Mamba architecture. Through comprehensive quantitative and qualitative analysis of real-world stock market data, including CSI, NASDAQ, and S\&P, we explore the ability of \name to effectively extract features from financial time series and dynamic relational graphs enriched by market feedback. Our analysis demonstrates its potential for accurate stock movement prediction across diverse market environments. Looking ahead, we will further explore the broader applications of the Mamba architecture in quantitative trading.


\clearpage
\balance
\bibliographystyle{ACM-Reference-Format}
\bibliography{sample-base}

\clearpage
\appendix

\section{Experiment Details}

\subsection{Datasets Details}\label{app:data}
For the four datasets (S\&P 500, NASDAQ 100, CSI300, CSI500), the train set is from 2018-01-01 to 2021-12-31, the validation set is from 2022-01-01 to 2022-12-31, and the test set is from 2023-01-01 to 2023-12-31. The statistics of each dataset is shown in \cref{tab:detaildataset}.

\renewcommand{\arraystretch}{0.5}
\begin{table}[htb]
\setlength{\tabcolsep}{6.5pt}
\caption{Detailed dataset descriptions.}
\resizebox{0.475\textwidth}{!}
{
\begin{tabular}{c|c|c|c|c}
\toprule

Dataset & Stocks (Nodes) & train set & validation set & test set \\

\midrule
CSI 300 & 285 & 943 & 242 & 242  \\
\midrule
CSI 500 & 450 & 943 & 242 & 242  \\
\midrule
S\&P 500 & 498 & 1008 & 251 & 249  \\
\midrule
NASDAQ 100 & 99 & 1008 & 251 & 249  \\

\bottomrule

\end{tabular}
}
\label{tab:detaildataset}
\end{table}

\subsection{Baseline Models}

We briefly describe the selected baseline models:

(1) Quantitative Investment Methods:

a. Classical Strategy:
\begin{itemize}
\item Buying-Loser-Selling-Winner (BLSW)~\cite{blsw}: which utilizes reversal strategies by identifying trends that are likely to reverse based on technical indicators and taking positions opposite to prevailing market direction.
\item Cross-Sectional Mean reversion (CSM)~\cite{csm}: whcih employs momentum strategies by identifying assets or securities with strong recent price trends and taking positions in the direction of those trends.
\end{itemize}

b. Deep Reinforcement Learning methods:
\begin{itemize}
\item AlphaStock~\cite{alphastock}: which is the first to offer an interpretable investment strategy using deep reinforcement attention networks. 
\item DeepPocket~\cite{deeppocket}: which consists of an autoencoder for feature extraction, a convolutional network to collect underlying information shared among financial instruments and an actor–critic RL agent. The source code is available at \url{https://github.com/MCCCSunny/DeepPocket}.
\end{itemize}

c. Deep Learning Methods:
\begin{itemize}
\item Transformer~\cite{attention} is a universal DL method featuring an Encoder-Decoder framework based on self-attention.
\item Mamba~\cite{mamba} is a state space model introducing a data-dependent selection mechanism that balances short-term and long-term dependencies.
\item FactorVAE~\cite{FactorVAE}: which integrates the dynamic factor model (DFM) with the variational autoencoder (VAE), and proposes a prior-posterior learning method which can approximate an optimal posterior factor model with future information. The source code is available at \url{https://github.com/ytliu74/FactorVAE}.
\item THGNN~\cite{thgnn}: which designs a temporal and heterogeneous graph neural network model to learn the dynamic relations among price movements in financial time series. The source code is available at \url{https://github.com/TongjiFinLab/THGNN}.
\item MambaStock~\cite{mambastock}: which mines historical stock market data with Mamba framework, eliminating the need for meticulous feature engineering or extensive preprocessing. The source code is available at \url{https://github.com/zshicode/MambaStock}.
\item CL~\cite{CL}: which predicts stock price movements by comparing textual and quantitative features of the current time interval against those of a prior time span for contrastive learning.
\item MASTER~\cite{master}: which models the momentary and cross-time stock correlation and leverages market information for automatic feature selection. The source code is available at \url{https://github.com/SJTU-DMTai/MASTER}.
\item CI-STHPAN~\cite{cisthpan}: which is a two-stage framework for stock selection, involving Transformer and HGAT based stock time series self-supervised pre-training and stock-ranking based downstream task fne-tuning. The source code is available at \url{https://github.com/Harryx2019/CI-STHPAN}.
\item VGNN~\cite{VGNN}: which is a decoupled graph learning framework for stock prediction with a tensor-based fusion module, a hybrid attention mechanism and a message-passing mechanism. The source code is available at \url{https://github.com/JiwenHuangFIC/VGNN}.
\end{itemize}

(2) Time Series Forecasting Methods:
\begin{itemize}
\item PatchTST~\cite{patchtst} is a Transformer-based model utilizing patching and channel independence technique. It also enable effective pre-training and transfer learning across datasets. The source code is available at \url{https://github.com/yuqinie98/PatchTST}.
\item iTransformer~\cite{itransformer} embeds each time series as variate tokens and is a fundamental backbone for time series forecasting. The source code is available at \url{https://github.com/thuml/iTransformer}.
\item TimeMixer~\cite{timemixer} is a Linear-based model enabling the combination of the multiscale information in both history extraction and future prediction phases. The source code is available at \url{hhttps://github.com/kwuking/TimeMixer}.
\item Crossformer~\cite{crossformer} is a Transformer-based model introducing the Dimension-Segment-Wise (DSW) embedding and Two-Stage Attention (TSA) layer to effectively capture cross-time and cross-dimension dependencies. The source code is available at \url{https://github.com/Thinklab-SJTU/Crossformer}.
\end{itemize}

\subsection{Metric Details}\label{app:metric}

We employ five widely used metrics to evaluate the overall performance of each method: Annual Return Ratio (ARR), Annual Volatility (AVol), Maximum Draw Down (MDD), Annual Sharpe Ratio (ASR) and Information Ratio (IR). The lower the absolute values of AVol and MDD, the higher the value of ARR, ASR, and IR, and the better the performance.
\begin{itemize}
    \item ARR measures the percentage increase or decrease in the value of an investment over the course of a year.
    \begin{equation}
        \text{ARR}=(1+\text{Total Return})^{\frac{1}{n}}-1
    \end{equation}
    \item AVol quantifies the volatility of an investment's return over the course of a year. $R_p$ denotes the daily return of the portfolio.
    \begin{equation}
        \text{AVol}=\sqrt{\text{Var}(R_p)}
    \end{equation}
    \item MDD represents the maximum drop from a peak to a trough in an investment's value.
    \begin{equation}
        \text{MDD}=-\text{max}\bigg(\frac{p_{peak}-p_{trough}}{p_{peak}}\bigg)
    \end{equation}
    \item ASR measures the risk-adjusted return of an investment over one year.
    \begin{equation}
        \text{ASR}=\frac{\text{ARR}}{\text{AVol}}
    \end{equation}
    \item IR measures the excess return of an investment relative to a benchmark adjusted for its volatility. $R_b$ is the daily return of the market index. 
    \begin{equation}
        \text{IR} = \frac{\text{mean}(R_p-R_b)}{\text{std}(R_p-R_b)}
    \end{equation} 
\end{itemize}

\subsection{Implementation Details}\label{app:imp}
Our experiment is trained on one NVIDIA V100 GPU, and all models are built using PyTorch~\cite{pytorch}. The training and validation sets are kept consistent for all baseline models. The number of GNN layers is $2$, the level in MLM is $2$, and the lookback horizon is $20$. The learning rate is $0.01$ for S\&P 500 and NASDAQ 100, and $0.03$ for CSI 300 and CSI 500. The weight of the hinge loss is set to $3.0$, while the weight of the GIB loss is set to $1.0$. We set $k$ to $9$ for the selected top-$k$ ranked stocks in the experiment.

\section{Extra Experimental Results}

Here we provide the extra results and analysis of experiments for \name.

\subsection{Additional Cases of Dynamic Interplay Between Stock Market Indices and Inter-Stock Correlations}

As shown in \cref{fig:corr&index}, the correlation between stocks becomes stronger when the market index falls and weaker when the market index rises. 
In particular, the relationship between stocks becomes very tight during several market downturns. 
For example, US stocks experienced a sharp pullback in the fourth quarter of 2018, driven by a combination of inflation concerns, interest rate hikes, a global economic slowdown and technical market factors. 
Similarly, from January to October 2022, inflationary pressures, post-pandemic supply chain disruptions and Federal Reserve policy shifts led to a prolonged decline in US equities.

\subsection{System Deployment Framework}
\begin{figure}[!ht]
    \centering
    \includegraphics[width=0.48\textwidth]{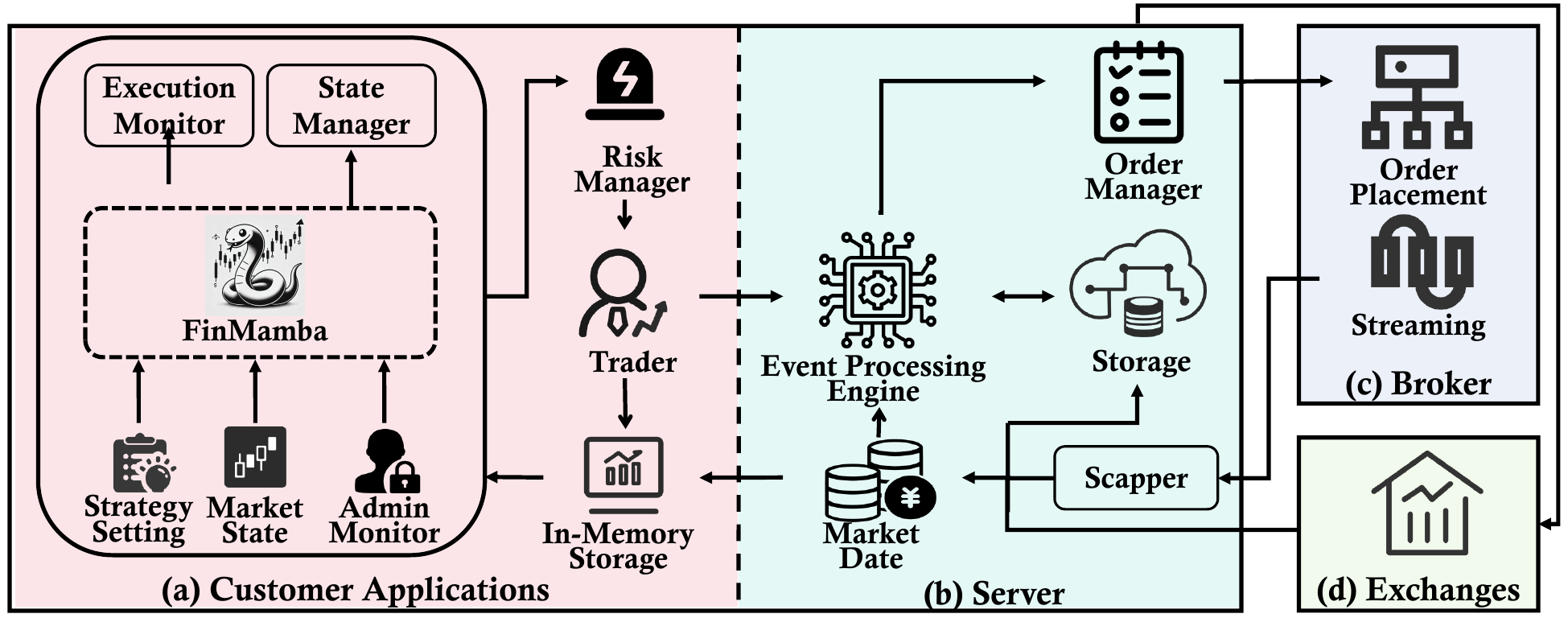}
    \caption{
         The system framework of our proposed \name
in quantitative trading scenarios.
    }
    \label{fig:deploy_framework}
\end{figure}

To validate our proposed \name in a realistic environment, we conduct an online deployment in the Chinese stock market.
The deployed architecture of \name is shown in \cref{fig:deploy_framework}. To efficiently adapt to changing market conditions, the \name is updated offline once a week. The newly updated model is then used to make online trading decisions throughout the following week.
In the live trading process, the server connects trading signals directly to the exchanges, with the event processing engine managing the flow of real-time trade orders. The market data system provides investors with up-to-date market information. The server database aggregates three types of data: live exchange data, historical data stored in memory, and real-time streaming data obtained from the broker's scraper service. All of this data is refined and transformed into actionable market information.
Trader applications subscribe to these market data, which are then stored in an in-memory engine before being passed to the \name trading decision agent.
Trading signals generated by \name are first validated by an independent risk management system before being passed to the event processing engine. Final trade orders are routed to the Order Manager, where they are encrypted using exchange-provided APIs before being sent back to the exchange for execution.

\subsection{Online Deployment}

\begin{figure}[!ht]
    \centering
    \includegraphics[width=0.48\textwidth]{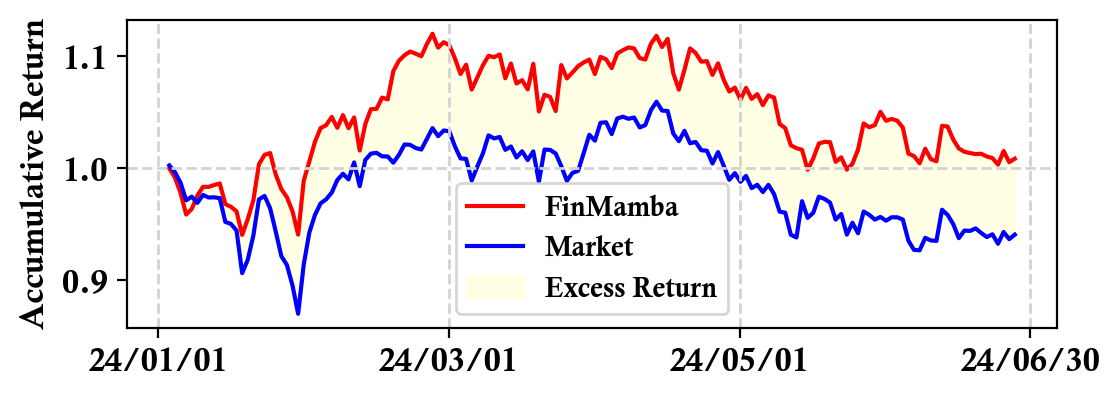}
    \caption{
         The performance of the strategy backtest.
    }
    \Description{..}
    \label{fig:deployment}
\end{figure}

To validate our model in a realistic environment, we conduct an online deployment based on the above framework in the Chinese stock market from 1 January 2024 to 30 June 2024.
Based on the predictions, we employ different strategies to trade within the first half hour of the next trading day’s opening.
The accumulative wealth of \name and the market return are depicted in \cref{fig:deployment}. Over half a year, all strategies in our model significantly outperform the market.

\end{document}